\documentclass[12pt,preprint]{aastex}

\newcommand{\myemail}{shogo@z.phys.nagoya-u.ac.jp}

\shorttitle{Distance to the Galactic Center}
\shortauthors{Nishiyama et al.}

\begin{document}

\title{The Distance to the Galactic Center\\
Derived
From Infrared Photometry of Bulge Red Clump Stars}

\author{Shogo Nishiyama\altaffilmark{1}, 
Tetsuya Nagata\altaffilmark{2},
Shuji Sato\altaffilmark{1},
Daisuke Kato\altaffilmark{1},
Takahiro Nagayama\altaffilmark{2},
Nobuhiko Kusakabe\altaffilmark{3}, 
Noriyuki Matsunaga\altaffilmark{4}, 
Takahiro Naoi\altaffilmark{5},
Koji Sugitani\altaffilmark{6}, and
Motohide Tamura\altaffilmark{7}
}

\altaffiltext{1}{Department of Astrophysics, Nagoya University, 
Nagoya 464-8602, Japan; \myemail}

\altaffiltext{2}{Department of Astronomy, Kyoto University, 
Kyoto 606-8502, Japan}

\altaffiltext{3}{Department of Astronomical Sciences,
Graduate University for Advanced Studies (Sokendai),
Mitaka, Tokyo 181-8588, Japan}

\altaffiltext{4}{Institute of Astronomy, School of Science, 
University of Tokyo, Mitaka, Tokyo, 181-0015, Japan}

\altaffiltext{5}{
Department of Infrared Astrophysics,
Institute of Space and Astronautical Science,
Japan Aerospace Exploration Agency,
Sagamihara, Kanagawa 229-8510, Japan
}

\altaffiltext{6}{Graduate School of Natural Sciences, Nagoya City University,
Nagoya, 467-8501, Japan}

\altaffiltext{7}{National Astronomical Observatory of Japan, 
Mitaka, Tokyo, 181-8588, Japan}

%------------------------------------------------------

\begin{abstract}

On the basis of the near infrared observations of bulge red clump stars
near the Galactic center,
we have determined the galactocentric distance to be
$R_0 = 7.52 \pm 0.10$ (stat) $\pm 0.35$ (sys) kpc.
We observed the red clump stars
at $\mid l \mid \la 1\fdg0$ and $0\fdg7 \la \mid b \mid \la 1\fdg0$
with the IRSF 1.4 m telescope and the SIRIUS camera
in the $H$ and $K_S$ bands.
After extinction and population corrections,
we obtained $(m - M)_0 = 14.38 \pm 0.03$ (stat) $\pm 0.10$ (sys).
The statistical error is
dominated by the uncertainty of the intrinsic local red clump stars' luminosity.
The systematic error is estimated to be $\pm 0.10$
including uncertainties in extinction and population correction,
zero-point of photometry, 
and the fitting of 
the luminosity function of the red clump stars.
Our result, $R_0 = 7.52$ kpc, is in excellent agreement
with the distance determined geometrically with the star
orbiting the massive black hole in the Galactic center.
The recent result based on the spatial distribution of globular clusters
is also consistent with our result.
In addition, our study exhibits that
the distance determination to the Galactic center 
with the red clump stars,
even if the error of the population correction is taken into account,
can achieve an uncertainty of about 5 percent,
which is almost the same level as that 
in recent geometrical determinations.

\end{abstract}


%------------------------------------------------------

\section{INTRODUCTION}
\label{sec:IntroDGC}

Since the review 
``The distance to the center of the Galaxy'' by \citet{Reid93},
the accuracy of the distance to the Galactic center (GC) from us, $R_0$,
has been improved especially in the primary measurements,
i.e., without a ``standard candle''.
\citet{Salim99} showed that
one can measure $R_0$ with an accuracy of a few percent
by observing the proper motions and radial velocities
of several stars near the super massive black hole (SMBH) at the GC
for 15$-$30 yr.
\citet{Eisen03,Eisen05} actually determined $R_0$
with a 4 percent uncertainty
by using one of the stars orbiting the SMBH.
Therefore, $R_0$ would be determined more precisely
than that estimated from the luminosity distance, i.e., the secondary measurements.

However, does the position of the SMBH coincide with the dynamical center
and/or the luminous center? 
Several lines of evidence for a displacement of the centroid of the Galactic bar
from the center of mass of our Galaxy exist
\citep[][and references therein]{Blitz94,Morris96},
and the off-centering of the stellar bar is predicted
from the simulation of the Milky Way model \citep[e.g.,][]{Fux99}
although the compact radio source Sgr A$^{*}$
is nearly at rest at the dynamical center of the Galaxy \citep{Reid04}.
Off-centered bars are commonly observed 
in external galaxies \citep[e.g.,][and references therein]{Levine98}.
The displacement would also reveal itself 
as a global tilt of the inner disk of the Galaxy.
A tilt of the triaxial bulge has been suggested \citep{Blitz91},
but the studies using  \textit{COBE} data sets
could not find statistically significant evidence for the tilt
\citep[e.g.,][]{Dwek95}.
The understanding of these effects is thus still 
in its earliest stages.

Red clump (RC) giants have been recently claimed to be
a very accurate distance indicator 
\citep[e.g.,][]{Demers94},
and a good tracer of the Galactic bar
\citep[e.g.,][]{Stanek97,Nishi05}.
RC stars are the equivalent of the horizontal-branch stars for 
a metal-rich population,
have narrow distributions in luminosity and color, 
and consequently occupy a distinct region in the color magnitude diagram (CMD).
The $Hipparcos$ catalog \citep{Hipp97} allows us
an exact calibration of an RC absolute magnitude;
therefore, RC stars can be used as a reliable standard candle.
The stellar density and the gravitational potential
of the Galaxy are well traced by 
low- and intermediate-mass, evolved stars
because they are dynamically relaxed,
and constitute the largest fraction of 
the total stellar mass.
Thus, by comparing the distance to the central stellar cluster
with that obtained by the distribution of bar components,
the existence of the displacement can be examined.

From photometric observations in infrared wavelength,
$R_0$ was already derived by \citet{Alves00} and \citet{Babu05}.
However, \citet{Alves00} used RC stars in Baade's window,
which is $\sim 4\degr \approx 560$ pc (if $R_0 = 8$ kpc) away from the GC,
and his error in $R_0$ is dominated by the small number
(about 200) of RC stars.
\citet{Babu05} observed ($l,b$) = ($0\fdg0, +1\fdg0$),
which is only 140 pc from the GC.
Since their purpose was to study the bar structure,
not to determine the distance to the GC,
they did not provide an exact value of $R_0$.
Therefore our determination, 
which employs RC stars 
at $\mid l \mid \leq 1\degr$ and $\mid b \mid \leq 1\degr$
and is almost free from
the statistical error due to abundant RC stars in the bulge,
is the most reliable ever studied with RC stars.

%------------------------------------------------------
\section{OBSERVATION AND DATA REDUCTION}
\label{sec:ObsDGC}

Our observations were conducted in 2002 March$-$July
and 2003 April$-$August
using the near-infrared camera SIRIUS
\citep[Simultaneous Infrared Imager for Unbiased Survey;][]{Nagas99, Nagay03} 
on board the IRSF (Infrared Survey Facility) telescope.
IRSF is a 1.4 m telescope constructed and operated 
by Nagoya University and SAAO (South African Astronomical Observatory)
at Sutherland, South Africa.
The SIRIUS camera
can provide $J$ (1.25 $\mu$m), $H$ (1.63 $\mu$m),
and $K_S$ (2.14 $\mu$m) images simultaneously,
with a field of view of 7\farcm7 $\times$ 7\farcm7
and a pixel scale of 0\farcs45.

\begin{figure}[h]
  \begin{center}
    \vspace{0.5cm}
    \epsscale{0.7}
    \rotatebox{-90}{
      \plotone{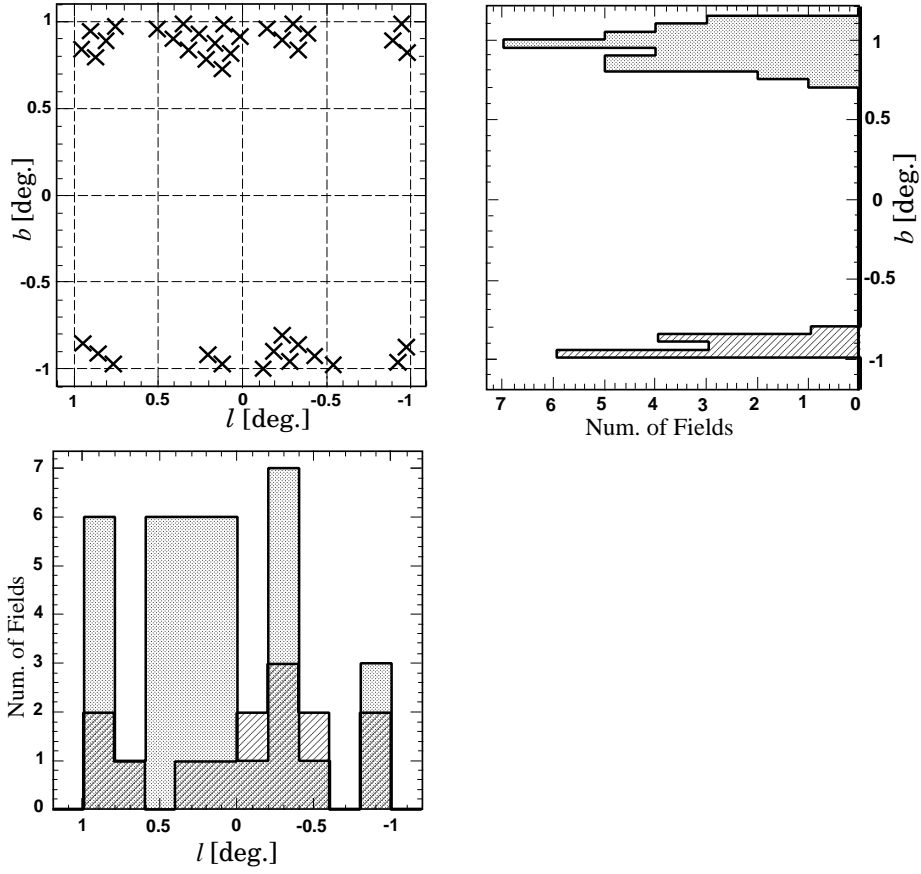}
    }
    \caption[]{
      \textit{Top left}: Fields used for data analysis,
      plotted in Galactic coordinates.
      \textit{Top right}: Distribution of the fields in Galactic latitude.
      \textit{Bottom left}: Distribution of the fields in Galactic longitude.
    }
    \label{fig:obsAreaGCDist}
  \end{center}
\end{figure}

About 100$\times$ 3 $(J,H,K_S)$
images were obtained 
over $\mid l \mid \la 1\fdg0$ and $\mid b \mid \la 1\fdg0$
(Fig. \ref{fig:obsAreaGCDist}).
We observed only on photometric nights,
and the typical seeing was 1\farcs1 FWHM in the $H$ band.
A single image comprises 10 dithered 5 s exposures.

Data reduction was carried out with 
the IRAF (Imaging Reduction and Analysis Facility)\footnote{
IRAF is distributed by the National Optical Astronomy
Observatory, which is operated by the Association of Universities for
Research in Astronomy, Inc., under cooperative agreement with
the National Science Foundation.}
software package.
Images were prereduced following the standard procedures
of near-infrared arrays 
(dark frame subtraction, flat-fielding, and sky subtraction).
Photometry, including point-spread function (PSF) fitting, was carried out 
with the DAOPHOT package \citep{Stetson87}.
We used the DAOFIND task to identify point sources,
and the sources were then utilized 
for PSF-fitting photometry in the ALLSTAR task.
About 20 sources were used to construct the PSF for each image.

Each image was calibrated with the standard star
9172 \citep{Persson98},
which was observed every hour in 2002
and every half-hour in 2003.
We assumed that star 9172 has $H=12.12$, and $K_S=12.03$
in the IRSF/SIRIUS system.
The average of the zero-point uncertainties
was about 0.03 mag in the two bands.
The averages of the 10 $\sigma$ limiting magnitudes were
$H=16.6$ and $K_S=15.6$.

To analyze the distribution of RC stars, 
we define the extinction-free magnitude
\begin{eqnarray}
  K_{H-K} \equiv K_S - \frac{A_{K_S}}{E(H-K_S)} \times 
   [ (H-K_S)-(H-K_S)_{0} ],
  \label{eq:DeredK}
\end{eqnarray}
where we use the reddening law 
${A_{K_S}}/{E(H-K_S)} = 1.44$ \citep{Nishi06},
and the intrinsic $H-K_S$ color of RC stars 
$(H-K_S)_{0} = 0.07$ \citep{Bonatto04}.
Thus, $K_{H-K}$ is defined so that 
if ${A_{K_S}}/{E(H-K_S)}$ is independent of location,
then $K_{H-K}$ is independent of
extinction for any particular star.

We construct $K_S$ vs. $H-K_S$ CMDs
and extract the stars in the region 
of the CMDs dominated by RC stars.
The extracted stars are used in turn
to make $K_{H-K}$ histograms 
(luminosity functions, see Fig. \ref{fig:KHKHist}).
The histograms have clear peaks
and are fitted with the sum of exponential and Gaussian functions
(\textit{thick curve} in Fig. \ref{fig:KHKHist}).
Owing to highly nonuniform interstellar extinction 
over the region observed,
the peaks of RC stars in the CMDs shift
from one line of sight to another
over the range $13.0 \lesssim K_S \lesssim 14.5$
although a dispersion of $K_{H-K}$ is small
(see \S \ref{subsec:R0disp}).
Since the mean $H$ and $K_S$ magnitudes of RC stars become too faint 
in highly reddened fields,
estimates of the peak magnitudes and the colors of RC stars
can be unreliable in such fields.
To avoid this problem, we use only those fields
in which the peak magnitude of RC stars is more than 1 mag brighter
than the 10$\sigma$ limiting magnitudes (Fig. \ref{fig:obsAreaGCDist}). 
In addition, we confirmed the completeness to be 85\% at $K_S = 15$
by adding artificial stars into the most crowded image.

%------------------------------------------------------
\section{RESULTS}
\label{sec:ResultsDGC}

\subsection{The Distance to the Galactic Center}

\begin{figure}[h]
 \begin{center}
  \epsscale{0.5}
  \rotatebox{-90}{
  \plotone{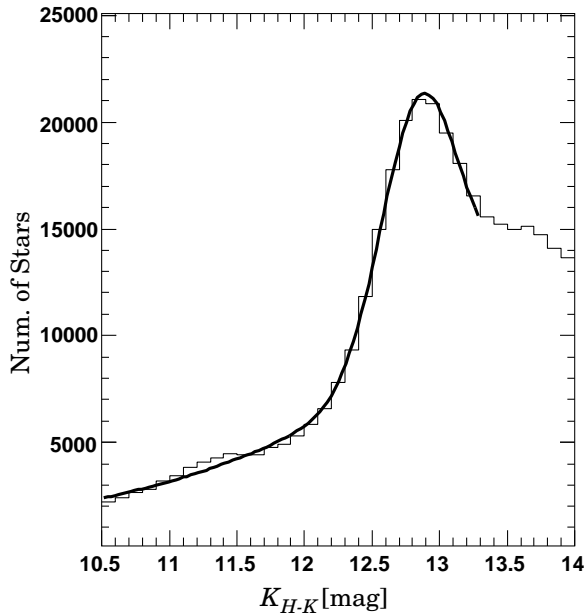}
  }
 \end{center}
  \caption{ Histogram of the dereddened $K_{H-K}$ magnitude
    for all the stars in the RC-dominated region
    at $\mid l \mid \la 1\fdg0$ and $0\fdg7 \la \mid b \mid \la 1\fdg0$.
    Exponential and Gaussian functions are used to fit 
    the histograms (\textit{thick curve}). 
  }
  \label{fig:KHKHist}
\end{figure}

By fitting the luminosity function of the extracted stars 
at $\mid l \mid \la 1\fdg0$ and $0\fdg7 \la \mid b \mid \la 1\fdg0$,
we obtained the center of the RC peak as $K_{H-K} = 12.855 \pm 0.005$
(Fig. \ref{fig:KHKHist}).
The distance modulus to the GC is given by
\begin{eqnarray}
  (m - M)_0 = K_{H-K} - M_{K_S} + \Delta M_{K},
  \label{eq:}
\end{eqnarray}
where $M_{K_S}$ is the absolute $K_S$ magnitude of local RC stars,
and $\Delta M_{K}$ is the population correction calculated 
from theoretical stellar evolution models.
Here we adopt 
$\Delta M_{K} = -0.07$ \citep[scaled solar metallicity,][]{Sala02}.
$M_{K_S}$ is estimated as
$M_{K_S} = M_{K} + \Delta M_{K_S-K}$,
where $M_{K}$ is the absolute $K$ magnitude
and $\Delta M_{K_S-K}$ is the difference 
between magnitudes in the $K_S$ and $K$ bands.
\citet{Alves02} obtained $M_K$ $= -1.60 \pm 0.03$ mag,
and \citet{Bonatto04} showed that $\Delta M_{K_S-K} \approx 0.01$,
and thus $M_{K_S} = -1.59$.
Hence, we obtain $(m - M)_0 = 14.38 \pm 0.03$ (stat),
which corresponds to $R_0 = 7.52 \pm 0.10$ (stat) kpc,
and the resulting statistical error is the sum of the squared errors.
Systematic errors are estimated below.

\subsection{Error Estimation}
\label{subsec:EstErDGC}

It is well known that
the statistical error in the determination of distances
using RC stars is very small \citep[e.g.,][]{Pac98}.
From the fitting shown in Fig. \ref{fig:KHKHist},
we obtain the center of the RC peak of $K_{H-K} = 12.855 \pm 0.005$.
As shown below, systematic errors are much larger than this,
and thus our statistical error is negligible in this determination.
However, the determination of the absolute magnitude of the local RC stars
has a statistical error of 0.03 \citep{Alves02},
and the total statistical error $\sigma_{\mathrm{stat}}$
in $(m-M)_0$ is therefore estimated 
to be $\sigma_{\mathrm{stat}} = 0.03$ mag.

One of the systematic errors arises from the fitting of the RC peak,
and thus we estimate the error by changing the fitting range
of the exponential and Gaussian functions.
The lower limit used in the fitting is changed from 10.5 to 12.0 in $K_{H-K}$,
and the upper limit is changed from 13.1 to 13.5.
When the upper limit is larger than 13.6,
the fitting fails because of the slope change around 13.6
(see, Fig. \ref{fig:KHKHist}).
The smallest and largest values of the peak center are 
12.824 and 12.885, respectively,
and thus we estimate that the uncertainty in the fitting is $\pm 0.03$ mag.

In the analysis of the interstellar extinction law,
\citet{Nishi06} showed that the law could change
even in the near infrared wavelengths;
at $\mid l \mid < 2\degr$ and $\mid b \mid  < 1\degr$,
$A_{K_S}/E_{H-K_S}$ changes by $\approx \pm 0.09$. %6\%$.  
Since the mean extinction correction is $\sim 0.6$ in $H - K_S$,
the systematic error in the extinction correction
is estimated to be $\pm 0.05$ mag.

A systematic error in our observation comes from 
the uncertainty of the zero-point determination 
of photometry.
From the analysis of the standard stars,
we obtained the mean zero-point uncertainty
to be about 0.04 mag in $K_{H-K}$.

The population correction $\Delta M_{K}$ was provided by
the analysis of the RC brightness 
with the theoretical stellar evolution model \citep{Sala02}.
We adopted $\Delta M_{K} = M_{K}^{\mathrm{local}} - M_{K}^{\mathrm{bulge}} = -0.07$
with solar metallicity.
The systematic error for the population correction
is difficult to estimate because
the bulge star formation rate and
age-metallicity relation should be involved.
[For the Large Magellanic Cloud, \citet{Sala03} estimated the systematic error.]
In this study, we adopt the 
systematic error for the correction $\pm 0.07$ 
\citep{Perciv03},
which is the 1 $\sigma$ error of the mean residual in $\Delta M_{K}$
between the model predicted and observed RC magnitudes.
Since \citet{Sala03} obtained the \textit{total} systematic error
in the distance modulus to the Large Magellanic Cloud as $^{+0.05}_{-0.06}$,
the systematic error $\pm 0.07$ in only $\Delta M_{K}$ might be overestimated.
Detailed analysis of the uncertainty and empirical tests
will provide us with more accurate $\Delta M_{K}$.

We can estimate the systematic error in system transformation
by comparing isochrones in different filter systems.
The isochrones presented in \citet{Girar02} are available online\footnote{
http://pleiadi.pd.astro.it}.
The difference in RC magnitudes and colors 
between the Johnson-Cousins-Glass system \citep{Bessell88}
and the 2MASS system\footnote{
http://www.ipac.caltech.edu/2mass/releases/second/doc/sec3\texttt{\symbol{`_}}1b1.html}
are $\approx 0.01$  in $K_S$
(also mentioned in Bonatto, Bica, \& Girardi, 2004) 
and $\approx 0.007$ in $H-K_S$, respectively.
The filter systems of 2MASS and SIRIUS are more similar than
those of 2MASS and the Johnson-Cousins-Glass system,
and thus the systematic error coming from the difference in the filter system
is smaller than 0.01, which is negligible in our analysis.

Since the fields we observed are very crowded,
stars might be blended and appear brighter than they really are.
To examine this effect,
we checked for the dependence of the RC peaks 
on star density using the 38 fields shown in
Fig. \ref{fig:obsAreaGCDist}
($\sim 300$ and $\sim 200$ stars arcmin$^{-2}$ 
[$ > 10 \sigma$, $K_S$ band]
for the most crowded and uncrowded regions, respectively),
and found no dependence.
Moreover,
we confirmed that the completeness is 85\% at $K_S = 15$.
The crowding effect is thus very small in this study.

\begin{table}[h]
 \begin{center}
  \caption{Systematic Error Budget}
  \vspace{0.1cm}
  \begin{tabular}[c]{lc}\hline \hline
    Error & Estimation (mag) \\ \hline
    fitting the RC peak & 0.03 \\
    extinction law & 0.05 \\
    zero-point uncertainty & 0.04 \\
    population correction & 0.07 \\
    system transformation & $<$ 0.01 \\
    \hline
  \end{tabular}
  \label{tab:SysError}
 \end{center}
\end{table}

The systematic errors in our analysis 
are summarized in Table \ref{tab:SysError}.
Combining these errors,
we obtained the total systematic error $\sigma_{\mathrm{sys}} = 0.10$.
Therefore, we obtain $(m - M)_0 = 14.38 \pm 0.03$ (stat) $\pm 0.10$ (sys),
which corresponds to $R_0 = 7.52 \pm 0.10$ (stat) $\pm 0.35$ (sys) kpc.

\subsection{$R_0$ Dispersion}
\label{subsec:R0disp}

The Galactic latitude and longitude distributions of 38 fields
we used in our analysis are shown in Fig. \ref{fig:obsAreaGCDist}.
The averages of longitude in positive and negative Galactic latitudes are
0.95 (0.11 rms uncertainty) and $-0.92$ (0.05 rms), and 
those of latitude in positive and negative Galactic longitudes are
$-0.16$ (0.52 rms) and 0.09 (0.58 rms), respectively.
Hence, the fields we observed are distributed nearly symmetrically around the GC,
which provides a good sample to examine the distribution of RC stars.

\begin{figure}[p]
 \begin{center}
  \vspace{0.5cm}
  \epsscale{0.9}
  \plotone{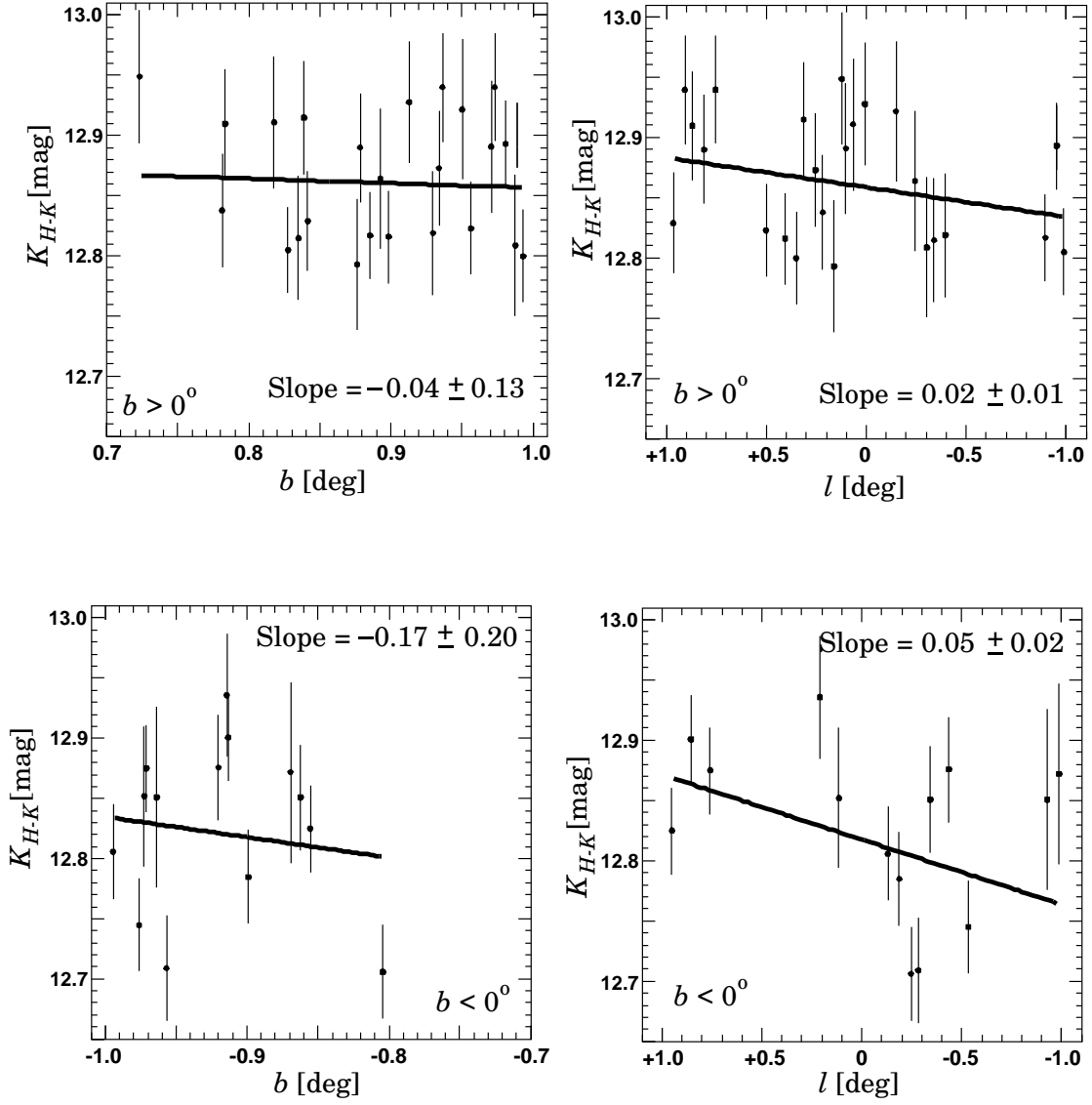}
  \caption{
  Plots of RC peaks in $K_{H-K}$
  as functions of Galactic latitude (\textit{left panels})
  and longitude (\textit{right panels}). 
  Top: Fields at $b > 0\degr$.
  Bottom: Fields at $b < 0\degr$.
  Straight lines are least-squares fits to the data points.
  }
  \label{fig:DisModlb}
 \end{center}
\end{figure}

Using the 38 fields shown in Fig. \ref{fig:obsAreaGCDist},
we obtained the peaks of RC stars in $K_{H-K}$ for each field,
and made plots as functions of Galactic latitude 
(\textit{left panels} in Fig. \ref{fig:DisModlb})
and longitude (\textit{right panels}). 
The top panels in Fig. \ref{fig:DisModlb} are 
the plots for the fields at $b > 0\degr$,
and the bottom panels are at $b < 0\degr$.
The straight lines in the plots are 
the least-squares fits to the data points.
The slopes given by the fits are
$-0.04 \pm 0.13 $ for $b > 0\degr$ and 
$-0.17 \pm 0.20 $ for $b < 0\degr$ as a function of Galactic latitude, and
$0.02 \pm 0.01 $ for $b > 0\degr$ and 
$0.05 \pm 0.02 $ for $b < 0\degr$ as a function of Galactic longitude.
These slopes suggest that the distribution
of RC stars along the lines of sight
is almost flat in our observed fields.

\begin{figure}[h]
 \begin{center}
  \vspace{0.5cm}
  \epsscale{0.6}
  \plotone{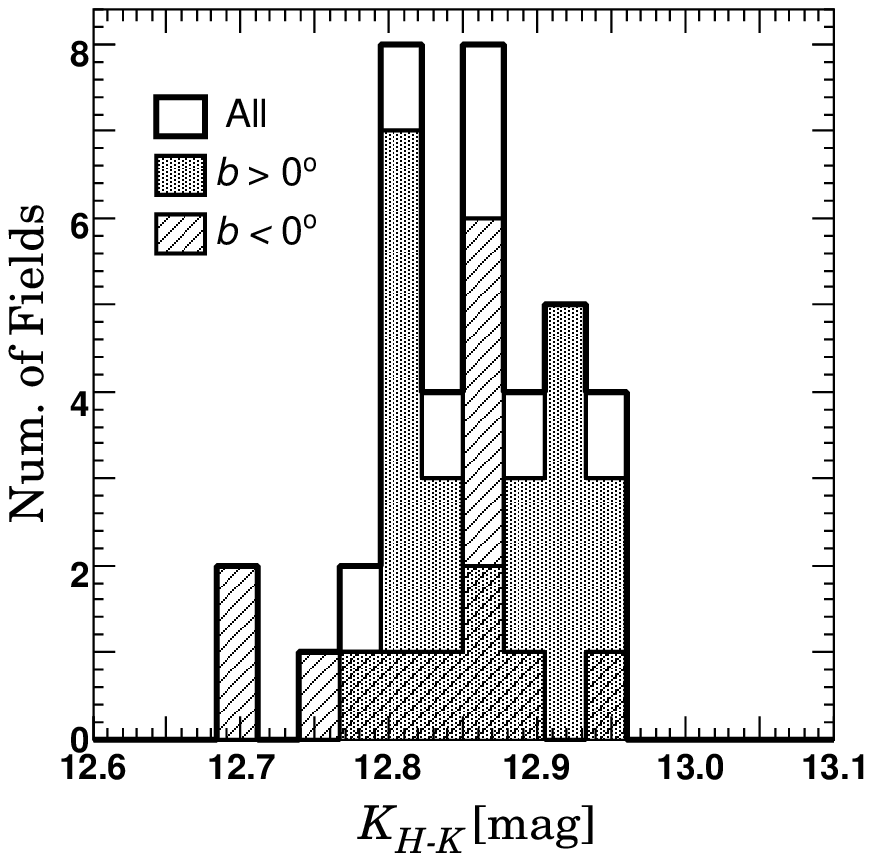}
    \caption{
      Histograms of RC peaks in $K_{H-K}$ 
      for all the fields (\textit{white histogram}),
      the fields at $b > 0\degr$ (\textit{dotted histogram})
      and those $b < 0\degr$ (\textit{hatched histogram}).
  The mean $K_{H-K}$ and rms are
      12.85 and 0.06 (all), 12.87 and 0.05 ($b > 0\degr$),
      and 12.83 and 0.07 ($b < 0\degr$), respectively.
    }
    \label{fig:DisModHist}
 \end{center}
\end{figure}

The RC peak distribution for all the fields 
is shown in Fig. \ref{fig:DisModHist} ({\it white histogram}).
From the histogram, we determine the mean $K_{H-K}$ to be 12.85
and the rms uncertainty of the distribution to be 0.06.
Because the statistical errors of the RC peak in $K_{H-K}$
are very small ($< 0.01$),
this uncertainty comes from the systematic uncertainties
in the fitting of the $K_{H-K}$ distributions (0.03; see Table \ref{tab:SysError}),
the extinction law (0.05), and the zero-point of standard stars (0.04). 
The sum of these squared errors is 0.07,
consistent with the rms uncertainty of 0.06 obtained above.
This consistency shows 
the validity of our estimate of the systematic errors.

The dotted and hatched histograms in Fig. \ref{fig:DisModHist} 
show the RC peak distributions in $K_{H-K}$ at $b > 0\degr$ and $b < 0\degr$, respectively.
The mean $K_{H-K}$ and uncertainty for $b > 0\degr$ are 12.87 and 0.05,
and those for $b < 0\degr$ are 12.83 and 0.07, respectively.
These also indicate that
the distances to bulge RC stars in positive and negative Galactic latitudes
are almost the same.

%------------------------------------------------------

\section{DISCUSSION}
\label{sec:DiscDGC}

We calibrated the absolute magnitude of RC stars
by using the theoretical prediction for the stars 
in Baade's window \citep{Sala02}.
Different populations could make a difference in the absolute magnitudes
between our observational fields and Baade's window.
However, the bulk of stellar populations in the bulge is old \citep[e.g.,][]{vanLoon03},
and the difference in age coming from 
the star formation histories employed in the models
affects $M_{K}$ much less
because the RC magnitude fades very slowly for old populations 
\citep[see also][]{Girar01}.
In addition, \citet{Ram00} examined a metallicity gradient
along the minor axis of the inner bulge 
($0\degr < b < -4\degr$ at $l \sim 0\degr$)
with the result that the gradient is negligible.
We therefore conclude that the RC absolute magnitudes 
in our fields and Baade's window are the same.

Our result $R_0 = 7.52 \pm 0.10$ (stat) $\pm 0.32$ (sys) kpc,
is in excellent agreement with $7.62 \pm 0.32$ kpc \citep{Eisen05},
which is determined geometrically with the star S2
orbiting the SMBH in the GC.
Recently, \citet{Bica06} provided a new estimate of $R_0$
based on symmetries of the spatial distribution of 116 globular clusters,
and obtained an average value of $R_0 = 7.2 \pm 0.3$ kpc.
This is smaller than those
derived by \citet{Eisen05} or adopted in recent reviews
($8.0 \pm 0.5$ kpc by Reid [1993], $7.9 \pm 0.2$ kpc by Nikiforov [2004]),
but is consistent with our result.
In \citet{Bica06}, numerous globular clusters are employed,
and the variation of total to selective absorption $R_V$ was taken into account;
thus the uncertainty in $R_0$ is a factor of $\sim 3$
smaller than previous ones using globular clusters.

Previously, $R_0$ was determined from the $K$ and $K_S$ band photometry
of RC stars \citep{Alves00,Babu05,Nishi05}.
Assuming $M_{K_S} = -1.61$, \citet{Babu05} obtained
$R_0 = 7.6 \pm 0.15$ kpc
without the population correction ($\Delta M_{K}$).
Because $\Delta M_{K}$ for the Galactic bulge makes 
the value of $R_0$ smaller than that without $\Delta M_{K}$,
the result of \citet{Babu05} suggests that
$R_0$ is significantly smaller than 8 kpc.
By contrast, \citet{Alves00} derived 
$R_0 = 8.24 \pm 0.42$ kpc, which is $\sim 2 \sigma$ larger than our result.
However, he did not adopt the population effect; with
$\Delta M_{K} = -0.07$, his distance becomes $R_0 = 7.98$ kpc.
The error in his distance modulus is dominated by a statistical one ($\pm 0.11$ mag)
because of the small number of RC stars in Baade's window.
When the statistical error is taken into account,
his distance is consistent with our result.
Assuming $(H-K_S)_{0} = 0.1$,
\citet{Nishi05} also derived $R_0 \approx 8.3$ kpc 
without the population correction.
When $(H-K_S)_{0} = 0.07$ and $\Delta M_{K} = -0.07$ are adopted,
their result is in good agreement with the current estimate.

Is the center of the RC distribution more distant
than 8 kpc (from the recent review by Reid 1993)
or 8.5 kpc (IAU 1985 standard; Kerr \& Lynden-Bell 1986)?
When $\Delta M_{K} = -0.07$ is applied,
the distance derived by \citet{Alves00} and \citet{Babu05}
with RC stars also become smaller than 8 kpc.
The star formation rate and age-metallicity relation of the Galactic bulge
is extremely complicated,
and theoretical prediction is troublesome.
In addition,
metallicity dependence of $\Delta M_{K}$ becomes considerable
at an age larger than $\sim 7$ Gyr \citep{Sala02},
and there are few old and metal rich clusters
to be compared with theoretical prediction
\citep{Groc02,Perciv03}.
However, to be $R_0 \approx 8.0$ kpc,
we require $\Delta M_{K} = +0.5$,
which is much larger and would be unacceptable from the theoretical model.
Therefore, we conclude that the center of the RC distribution is
at most nearer than 8 kpc.

Some observations suggest 
a displacement of the centroid of a bar 
from the central stellar cluster in our Galaxy.
A mode of $(m,l) = (1,1)$, 
where a star completes $l$ orbits
while executing $m$ epicycle oscillations as seen in the rotating frame,
could make the displacement \citep{Sellwood93}.
The mode also reveals itself as a global tilt 
of the inner disk of the Galaxy.
Several pieces of evidence exist 
for the mode of $(m,l) = (1,1)$, such as 
the pronounced longitudinal asymmetry of molecular line emissions,
and the tilt of the plane defined by gas x$_1$ orbits
with respect to the large-scale Galactic plane 
\citep[][and references therein]{Blitz94,Morris96}.
The studies of bulge structure \citep[e.g.,][]{Stanek97,Nishi05}
showed that the RC star is a good tracer of 
the bar structure in the central region of our Galaxy.
Thus, the mode of $(m,l) = (1,1)$ could make a difference 
in the distances determined from the distribution of RC stars
and from the stars orbiting the SMBH at the center of our Galaxy.
The good agreement between our result and that of \citet{Eisen05}
indicates that the displacement along the line of sight
is negligible with an uncertainty of a few hundred pc.

%------------------------------------------------------
\section{CONCLUSION}
\label{sec:ConcDGC}

Using bulge red clump stars,
the distance to the Galactic center is determined.
Our distance modulus is estimated to be
$(m - M)_0 = 14.38 \pm 0.03$ (stat) $\pm 0.10$ (sys),
which corresponds to the distance
$R_0 = 7.52 \pm 0.10$ (stat) $\pm 0.35$ (sys) kpc. 
The statistical error is dominated by 
that of the local red clump magnitude used for calibration.
The main sources of our systematic error are
from uncertainties in extinction correction and
population correction
calculated from theoretical stellar evolution models.
Our analysis shows that
the distances to the red clump stars are almost the same
in the range of
$\mid l \mid \la 1\fdg0$ and $0\fdg7 \la \mid b \mid \la 1\fdg0$.

\acknowledgements

We would like to thank the IRSF/SIRIUS group for their helpful comments.
We also thank the staff at SAAO for their support during our observations.
The IRSF/SIRIUS project was initiated and supported by Nagoya
University, the National Astronomical Observatory of Japan
and the University of Tokyo in collaboration with 
the South African Astronomical Observatory under 
Grants-in-Aid for Scientific Research
No.10147207, No.10147214, No.13573001, and No.15340061
of the Ministry of Education,
Culture, Sports, Science and Technology (MEXT) of Japan.
This work was also supported in part 
by the Grants-in-Aid for the 21st Century 
COE ``The Origin of the Universe and Matter: 
Physical Elucidation of the Cosmic History'' 
and ``Center for Diversity and Universality in Physics''
from the MEXT of Japan.

%--------------------------------------------------------

\end{document}